# Effect of low-temperature annealing on the electronic- and band-structure of (Ga,Mn)As epitaxial layers

O Yastrubchak [1,*], T Andrearczyk [2], J Z Domagala [2], J Sadowski [2,3], L Gluba [1], J Żuk [1] and T Wosinski [2]

[1] Institute of Physics, Maria Curie-Sklodowska University in Lublin, 20-031 Lublin, Poland

[2] Institute of Physics, Polish Academy of Sciences, 02-668 Warsaw, Poland

[3] MAX-IV Laboratory, Lund University, 221 00 Lund, Sweden

*E-mail: yastrub@hektor.umcs.lublin.pl

**Abstract**

The effect of outdiffusion of Mn interstitials from (Ga,Mn)As epitaxial layers, caused by post-growth low-temperature annealing, on their electronic- and band-structure properties has been investigated by modulation photoreflectance (PR) spectroscopy. The annealing-induced changes in structural and magnetic properties of the layers were examined with high-resolution X-ray diffractometry and SQUID magnetometery, respectively. They confirmed an outdiffusion of Mn interstitials from the layers and an enhancement in their hole concentration, which were more efficient for the layer covered with a Sb cap acting as a sink for diffusing Mn interstitials. The PR results revealing a decrease in the band-gap-transition energy in the as-grown (Ga,Mn)As layers, with respect to that in the reference GaAs one, are interpreted by assuming a merging of the Mn-related impurity band with the host GaAs valence band. On the other hand, an increase in the band-gap-transition energy in the annealed (Ga,Mn)As layers is interpreted as a result of the Moss-Burstein shift of the absorption edge due to the Fermi level location within the valence band, determined by the enhanced free-hole concentration. The experimental results are consistent with the valence-band origin of mobile holes mediating ferromagnetic ordering in (Ga,Mn)As, in agreement with the Zener model for ferromagnetic semiconductors.

# 1. Introduction

The ternary III-V semiconductor (Ga,Mn)As, combining semiconducting properties with magnetism, has become a representative diluted ferromagnetic semiconductor giving rise to a possible integration of electronic and magnetoelectronic devices prospective for future spintronic applications. Homogeneous (Ga,Mn)As layers containing up to about 20% of Mn atoms can be grown by the low-temperature (150–250°C) molecular-beam epitaxy (LT-MBE) technique [1–3]. When intentionally undoped, the layers are of *p*-type where Mn atoms substituting the Ga lattice atoms, $Mn_{Ga}$, supply both mobile holes and magnetic moments. Below the magnetic Curie temperature, $T_C$, the layers become ferromagnetic due to the hole-mediated ordering of substitutional $Mn^{2+}$ ions, with half-filled 3*d* shell and S = 5/2 spin moment, which has been successfully explained within the frames of the *p-d* Zener model for ferromagnetic semiconductors just over a decade ago [4]. However, theoretical understanding of the band structure and the underlying ferromagnetic interactions are still under debate [5,6].

Although the highest $T_C$ in the as-grown (Ga,Mn)As layers remains so far at 110°C [1, 3], it has been increased, as a result of post-growth low-temperature (180–250°C) annealing treatments, up to 185°C in thin epitaxial layers [7, 8] or even up to 200°C in (Ga,Mn)As layers subjected to nanostructurization [9]. The main effect of annealing, leading to the efficient increase in both the hole density and $T_C$, is outdiffusion of Mn interstitials, $Mn_I$, from the (Ga,Mn)As layers. According to first principles calculations [10] and channelling experiments [11] Mn interstitials act as double donors in GaAs; i.e. one $Mn_I$ atom compensates for the holes created by two substitutional $Mn_{Ga}$ atoms, thus resulting in effective reduction of the hole concentration in the layers and, in turn, in decreasing the ferromagnetic order as follows from the Zener model of carrier-mediated ferromagnetism. Moreover, the highly mobile positively charged $Mn_I$ can be immobilized by occupying interstitial sites adjacent to negatively charged substitutional $Mn_{Ga}$ acceptors, forming thus $Mn_I$–$Mn_{Ga}$ pairs. Such pairs become antiferromagnetically ordered, as theoretically predicted [12] and experimentally confirmed [13], and lead to further suppression of the total ferromagnetic moment of $Mn_{Ga}$ ions in (Ga,Mn)As. Adell et al. [14] have shown that deposition of a thin amorphous As film on the (Ga,Mn)As surface results in much shorter time and/or lower temperature of the desired post-growth annealing, where As capping acts as an efficient sink for diffusing Mn interstitials. Those authors obtained also similar results by using an amorphous Sb cap [14]. It is expected that further optimization of the MBE-growth conditions and the post-growth annealing will succeed in obtaining (Ga,Mn)As layers, with about 10% Mn content on substitutional sites, showing room-temperature ferromagnetism [15].

Two alternative models of the band structure in (Ga,Mn)As are predominant now. The first one, the Zener model, assumes mobile holes residing in the valence band of GaAs and the Fermi level position determined by the concentration of valence-band holes [4, 16]. The second one involves persistence of the narrow, Mn-related, impurity band even in highly Mn-doped (Ga,Mn)As with metallic conduction. In this model the Fermi level exists in the impurity band within the band gap and the mobile holes retain the impurity band character [17–19].

Our recent results, based on modulation photoreflectance (PR) spectroscopy investigations [20], are in favour of the Zener model. From analysis of the PR spectra obtained for (Ga,Mn)As layers with various Mn contents we revealed significant decrease in the band-gap-transition energy with increasing Mn content. We interpreted the results in terms of a disordered valence band, extended within the band gap, formed, in highly Mn-doped (Ga,Mn)As, as a result of merging the Mn-related impurity band with the host GaAs valence band, in agreement with the band structure proposed by Jungwirth et al. [21, 22]. The disordered character of the valence band may account for the observed very low mobility of itinerant holes in ferromagnetic (Ga,Mn)As layers. Our experimental results are also consistent with recent first-principles calculations of the (Ga,Mn)As band-gap energy [23] and the very recent experimental results of hard X-ray angle-resolved photoemission spectroscopy [24].

In the present paper we have applied modulation PR spectroscopy to study the band-structure evolution in (Ga,Mn)As eptaxial layers with 6% of Mn content under the low-temperature annealing treatment and the influence of Sb capping of the layer surface on the annealing-induced effects. Impact of the annealing treatments on structural and magnetic properties of the layers has been examined with high-resolution X-ray diffractometry (XRD) and superconducting quantum interference device (SQUID) magnetometery, respectively.

## 2. Experimental

We investigated two (Ga,Mn)As layers, with 6% Mn content and 100 nm thickness, grown by means of the LT-MBE technique at a temperature of 230°C on semi-insulating (001)-oriented GaAs substrates. After the growth one of the (Ga,Mn)As layers was covered, in the MBE chamber, by 10 monolayers of amorphous Sb cap to make the annealing procedure more efficient. This layer is labelled as (Ga,Mn)As/Sb. In addition, we investigated, as reference LT-GaAs layers, two undoped GaAs layers, of the thickness of 230 nm and 800 nm, grown on GaAs by LT-MBE under the same conditions as the (Ga,Mn)As layers. Both the Mn composition and the layer thickness were verified during growth by the reflection high-energy electron diffraction (RHEED) intensity oscillations [25]. After the growth, the wafers containing (Ga,Mn)As layers were cleaved into two parts. One part of each wafer was subjected to low-temperature annealing treatment performed in air at the temperature of 180ºC during 60 h.

The structural properties of the investigated layers were revealed from analysis of the results of high-resolution XRD measurements performed at room temperature with the X-ray diffractometer equipped with a parabolic X-ray mirror and four-bounce Ge 220 monochromator at the incident beam and a three-bounce Ge analyzer at the diffracted beam. Misfit strain in the layers was investigated using reciprocal lattice mapping and rocking curve techniques for both the symmetric 004 and asymmetric 224 reflections of Cu K$\alpha_1$ radiation. The magnetic properties of the (Ga,Mn)As layers, both the as-grown and the annealed ones, were characterized with a SQUID magnetometer in the temperature range from 5 K to 150 K. Modulation PR measurements of all the layers were performed at room temperature using a semiconductor laser operating at the 532 nm wavelength and a nominal power of 50 mW, as a pump-beam source, and a 250 W halogen lamp coupled to a monochromator, as a probe-beam source. The PR signal was detected by a Si photodiode. The chopping frequency of the pump beam was 70 Hz and the nominal spot size of the pump and probe beams at the sample surface were 2 mm in diameter. Owing to the derivative nature of modulation PR spectroscopy it allows for accurate determination of the optical-transition energies even at room temperature.

## 3. Structural and magnetic characterization

High-resolution XRD measurements performed for the 230-nm thick LT-GaAs layer and the (Ga,Mn)As layers have shown that all the epitaxial layers were grown pseudomorphically on GaAs substrate under compressive misfit strain. The layers exhibited a high structural perfection, as proved by clear X-ray interference fringes revealed for the 004 Bragg reflections of all the layers, as shown in figure 1a and b. The layer thicknesses calculated from the angular spacing of the fringes correspond very well to their thicknesses determined from the growth parameters. Diffraction peaks corresponding to the LT-GaAs and (Ga,Mn)As layers in figure 1 shift to smaller angles, with respect to that of the GaAs substrate, as a result of larger lattice parameters. In LT-GaAs it is caused by incorporation of a large amount of about 1% excess arsenic, mainly in form of arsenic antisites, $As_{Ga}$ [26]. The lattice parameter in (Ga,Mn)As becomes more increased as a result of additional incorporation of the $Mn_{Ga}$ and $Mn_I$ atoms in the crystal lattice [27]. Angular positions of the diffraction peaks corresponding to epitaxial layers were used to calculate the perpendicular lattice parameters, $c$, and the relaxed lattice parameters, $a_{rel}$, [28] (assuming the (Ga,Mn)As elasticity constants to be the same as for GaAs). The results are listed in Table 1. The lattice unit of the layers changes with increasing lattice mismatch from the zinc-blende cubic structure to the tetragonal structure with the perpendicular lattice parameter larger than the lateral one, equal to the GaAs substrate lattice parameter, which is $a_{sub}$ = 5.65349 Å in our XRD experiments. The annealing treatment applied to both the (Ga,Mn)As layers resulted in a distinct increase in angular positions of their diffraction peaks and a decrease in their lattice parameters, shown in Table 1, which is in agreement with the expected outdiffusion of Mn interstitials from the (Ga,Mn)As layers [27].

The results of SQUID magnetometry applied to the (Ga,Mn)As layers are presented in figure 2a and b. They show that the layers exhibit an in-plane easy axis of magnetization, characteristic of compressively strained (Ga,Mn)As layers, and well defined hysteresis loops in their magnetization vs. magnetic field dependence shown in the insets in figure 2. The as-grown layers displayed the $T_C$ values of 40 K and 50 K. The higher $T_C$ value obtained for the (Ga,Mn)As/Sb layer (figure 2b) with respect to that for the (Ga,Mn)As one (figure 2a) suggests a little larger Mn content in the former layer, which is corroborated by slightly larger values of the lattice parameters for that layer; c.f. Table 1. As a result of the annealing treatment the Curie temperatures of both the layers increased significantly. As expected, the annealing was much more efficient in the case of the (Ga,Mn)As/Sb layer, resulting in its $T_C$ equal to 115 K, much larger than that of 73 K for the annealed (Ga,Mn)As layer. Formation of MnSb on top of the (Ga,Mn)As/Sb layer during the annealing should be excluded as a possible contribution to the increased Curie temperature in that layer since $T_C$ of small MnSb precipitates is as high as about 600 K, exceeding even $T_C$ = 585 K for the bulk MnSb, as recently shown by Kochura et al. [29]. The $T_C$ values for all the (Ga,Mn)As layers are listed in Table 1. The annealing treatment resulted, additionally, in a significant decrease in the coecivity of the (Ga,Mn)As layers and an increase in their remnant magnetization, as shown in the insets in figure 2a and b. The coercive field of 100 Oe in the two as-grown layers decreased, as a result of annealing, to 50 Oe in the (Ga,Mn)As layer and to 25 Oe in the (Ga,Mn)As/Sb one. This coercivity decrease mainly results from an increase in the hole concentration in the annealed layers [30].

## 4. Photoreflectance results and analysis

The modulation PR spectra measured in the photon-energy range from 1.35 to 1.65 eV for the LT-GaAs and (Ga,Mn)As epitaxial layers are presented in figure 3a and b. The PR signal is associated with the electric field caused by the separation of photogenerated charge carriers. As a result of screening of this electric field by free carriers the intensity of measured PR signal strongly depends on free carrier concentration in the investigated layers. The spectra presented in figure 3 have been normalized to the same intensity. All the experimental spectra reveal a complex, modulated structure containing a differentiated peak corresponding to the interband optical transition and the electric-field-induced Franz-Keldysh oscillations (FKOs) [31] at energies above the fundamental absorption edge, which are marked with numbers in figure 3a and b. Moreover, the PR spectra for all the (Ga,Mn)As layers and for the 230-nm thick LT-GaAs layer exhibit an additional feature, at the photon energy of about 1.42 eV, denoted with capital A in figure 3a. The nature of such a feature, appearing on the low-energy side of the band-gap transition in PR spectra, was investigated in details by Sydor et al. [32] for differently doped GaAs epitaxial layers of various thicknesses grown by MBE on semi-insulating GaAs substrate. This feature was interpreted as a contribution to the PR signal from the layer–substrate interface region. The modulation mechanism of the feature results from the thermal excitation of impurities or traps at the interface and their momentary refilling by the laser-injected carriers [32]. Lack of the feature A in our PR spectrum for the thicker, 800-nm thick, LT-GaAs layer, shown in figure 3a, strongly supports the above interpretation of its origin.

Because of the excitonic nature of the room-temperature PR signal, the measured energies for interband transitions in GaAs are smaller than the nominal band-gap energies by approximately the excitonic binding energy. Detailed numerical analysis of such complex PR spectra consists in the full-line-shape fitting of the spectra by using Airy and Aspnes third-derivative line-shape functions [33]. In our previous paper [20] we have shown that by employing a simplified analysis of the PR spectra for (Ga,Mn)As layers with various Mn contents, which consists in an analysis of the energy positions of FKOs extrema, we have obtained similar changes of the interband transition energies, while increasing Mn content in the layers, to those obtained from the full-line-shape fitting of the PR spectra.

In the present paper we confine ourselves to this simplified analysis of our PR spectra. This analysis, which takes the asymptotic expressions of the Airy functions and their derivatives [34], results in a linear dependence of the energy positions of FKOs extrema $E_m$ on their effective index, defined as $F_m = [3\pi(m-1/2)4]^{2/3}$, according to the relation [32]:

$$E_m = E_G + \hbar\theta F_m, \tag{1}$$

where $m$ is the extremum number, marked in figure 3a and b, $E_G$ is the interband transition energy and

$$\hbar\theta = \left(\frac{e^2\hbar^2 F^2}{2\mu}\right)^{1/3} \tag{2}$$

is the electro-optic energy, where $e$ and $\hbar$ have their usual meanings, $F$ is the electric field and $\mu$ is the interband reduced mass of the electron-hole pair. The dependences of $E_m$ versus $F_m$ for the investigated LT-GaAs and (Ga,Mn)As layers are plotted in figure 4. The values of $E_G$ and $\hbar\theta$, obtained from the intersection with ordinate and from the slope of this linear dependence, respectively, are listed in Table 1.

For the two investigated LT-GaAs layers, which differ considerably in their thicknesses, we have obtained very similar $E_G$ values but markedly different electro-optic energies $\hbar\theta$, resulting from different slopes of the corresponding dependences shown in figure 4. The smaller value of $\hbar\theta$ revealed for the thicker LT-GaAs layer is presumably caused by much lower electric field in this layer; cf. formula (2). High concentration of the $As_{Ga}$ defects in LT-GaAs results in the Fermi level position near the middle of the band gap [26, 35], determined by partial ionization of the $As_{Ga}$-related double donors by residual acceptors. However, owing to much larger dipole-transition strength for band-to-band optical transitions, than that for defect-to-band transitions, electronic transitions from the top of the valence band to the conduction band are predominant in the modulation PR spectra measured for LT-GaAs.

On the other hand, the plotted results for the two as-grown layers of (Ga,Mn)As and (Ga,Mn)As/Sb are very similar and almost indistinguishable in figure 4. They display the same value of $E_G = 1.408$ eV, which is red-shifted with respect to that of the LT-GaAs layers. This result is in qualitative agreement with our previous results [20] and with the proposed interpretation assuming a merging the Mn-related impurity band with the host GaAs valence band. The annealing treatment of the (Ga,Mn)As layers resulted in a blue-shift of the interband transition energy, which was distinctly stronger for the layer covered with the Sb cap. Taking into account that the main effect of the annealing treatment was an increase in the hole concentration in the layers, as concluded from our SQUID magnetometry results, we interpret this blue-shift as a result of the Moss-Burstein shift of the absorption edge [36] due to the Fermi level position, determined by the free-hole concentration, within the valence band. In line with this interpretation the higher hole concentration in the annealed (Ga,Mn)As/Sb layer, with respect to that in the (Ga,Mn)As one, results in stronger blue-shift of the interband transition energy.

## 5. Summary

We have studied the effect of out-diffusion of Mn interstitials from (Ga,Mn)As epitaxial layers, caused by low-temperature annealing, on their electronic- and band-structure properties by applying modulation photoreflectance spectroscopy. Structural and magnetic properties of the layers were characterized with high-resolution X-ray diffractometry and SQUID magnetometery, respectively. Two (Ga,Mn)As layers with 6% Mn content and two reference LT-GaAs layers grown by the LT-MBE technique were investigated. One of the (Ga,Mn)As layers was covered with a thin cap of amorphous Sb after the growth to make the annealing treatment more effective. The (Ga,Mn)As layers were subjected to the annealing treatment at 180ºC for 60 h in air. The results of XRD characterization revealed a decrease in lattice parameters of the (Ga,Mn)As layers caused by the annealing, confirming the efficient outdiffusion of Mn interstitials from the layers. SQUID magnetometry measurements demonstrated a significant increase in the Curie temperature and remnant magnetization and decrease in the coercivity in the (Ga,Mn)As layers, as a result of the annealing, suggesting an enhancement of the hole concentration in the layers. Both the increase in $T_C$ and decrease in the coercivity in the annealed (Ga,Mn)As layers were much more significant for the (Ga,Mn)As/Sb layer, testifying to efficiency of the Sb cap as a sink for diffusing Mn interstitials.

The modulation PR spectra for all the layers investigated revealed clear Franz-Keldysh oscillations at photon energies above the fundamental absorption edge. Thorough analysis of the periods of FKOs enabled determination of the evolution of the optical transition energy in the (Ga,Mn)As layers resulting from the annealing treatment. Decrease in the band-gap-transition energy in the as-grown (Ga,Mn)As layers, with respect to that in the reference LT-GaAs layers, is interpreted by assuming a merging of the Mn-related impurity band with the host GaAs valence band resulting in electronic transitions from the top of this disordered valence band to the conduction band. On the other hand, an increase in the band-gap-transition energy in the (Ga,Mn)As layers subjected to annealing is interpreted as a result of the Moss-Burstein shift of the absorption edge due to the Fermi level location within the disordered valence band, determined by the enhanced free-hole concentration. The experimental results are consistent with the valence-band origin of mobile holes, which mediate ferromagnetic ordering in (Ga,Mn)As, in accordance with the *p-d* Zener model for ferromagnetic semiconductors.

## Acknowledgments

O Y acknowledges financial support from the Foundation for Polish Science under Grant POMOST/2010-2/12 sponsored by the European Regional Development Fund, National Cohesion Strategy: Innovative Economy. This work was also supported by the Polish Ministry of Science and Higher Education under Grant No. N N202 129339. The MBE project at MAX-IV Laboratory is supported by the Swedish Research Council (VR).

# References

[1] Matsukura F, Ohno H, Shen A and Sugawara Y 1998 *Phys. Rev. B* **57** R195205

[2] Chiba D, Nishitani Y, Matsukura F and Ohno H 2007 *Appl. Phys. Lett.* **90** 122503

[3] Mack S, Myers R C, Heron T, Gossard A C and Awschalom D D 2008 *Appl. Phys. Lett.* **92** 192502

[4] Dietl T, Ohno H and Matsukura F 2001 *Phys. Rev. B* **63** 195205

[5] Richardella A, Roushan P, Mack S, Zhou B, Huse D A, Awschalom D D and Yazdani A 2010 *Science* **327** 665

[6] Samarth N 2012 *Nature Mater.* **11** 360

[7] Novak V, Olejnik K, Wunderlich J, Cukr M, Vyborny K, Rushforth A W, Edmonds K W, Campion R P, Gallagher B L, Sinova J and Jungwirth T 2008 *Phys. Rev. Lett.* **101** 077201

[8] Wang M, Campion R P, Rushforth A W, Edmonds K W, Foxon C T and Gallagher B L 2008 *Appl. Phys. Lett.* **93** 132103

[9] Chen L, Yang X, Yang F, Zhao J, Misuraca J, Xiong P and von Molnar S 2011 *Nano Lett.* **11** 2584

[10] Máca F and Mašek J 2002 *Phys. Rev. B* **65** 235209

[11] Yu K M, Walukiewicz W, Wojtowicz T, Kuryliszyn I, Liu X, Sasaki Y and Furdyna J K 2002 *Phys. Rev. B* **65** 201303(R)

[12] Blinowski J and Kacman P 2003 *Phys. Rev. B* **67** 121204(R)

[13] Edmonds K W, Farley N R S, Johal TK, van der Laan G, Campion R P, Gallagher B L and Foxon C T 2005 *Phys. Rev. B* **71** 064418

[14] Adell M, Ilver L, Kanski J, Stanciu V, Svedlindh P, Sadowski J, Domagała J Z, Terki F, Hernandez C and Charar S 2005 *Appl. Phys. Lett.* **86** 112501

[15] Jungwirth T, Wang K Y, Mašek J, Edmonds K W, König J, Sinova J, Polini M, Goncharuk N A, MacDonald A H, Sawicki M, Rushforth A W, Campion R P, Zhao L X, Foxon C T and Gallagher B L 2005 *Phys. Rev. B* **72** 165204

[16] Jungwirth T, Sinova J, Masek J, Kucera J and MacDonald A H 2006 *Rev. Mod. Phys.* **78** 809

[17] Berciu M and Bhatt R N 2004 *Phys. Rev. B* **69** 045202

[18] Burch K S, Shrekenhamer D B, Singley E J, Stephens J, Sheu B L, Kawakami R K, Schiffer P, Samarth N, Awschalom D D and Basov D N 2006 *Phys. Rev. Lett.* **97** 087208

[19] Dobrowolska M, Tivakornsasisthorn K, Liu X, Furdyna J K, Berciu M, Yu K M and Walukiewicz W 2012 *Nature Mater.* **11** 444

[20] Yastrubchak O, Żuk J, Krzyżanowska H, Domagala J Z, Andrearczyk T, Sadowski J and Wosinski T 2011 *Phys. Rev. B* **83** 245201

[21] Jungwirth T, Sinova J, MacDonald A H, Gallagher B L, Novák V, Edmonds K W, Rushforth A W, Campion R P, Foxon C T, Eaves L, Olejník E, Mašek J, Yang S-R E, Wunderlich J, Gould C, Molenkamp L W, Dietl T and Ohno H 2007 *Phys. Rev. B* **76** 125206

[22] Jungwirth T, Horodyska P, Tesarova N, Nemec P, Surbt J, Maly P, Kuzel P, Kadlec C, Masek J, Nemec I, Orlita M, Novák V, Olejník K, Soban Z, Vasek P, Svoboda P and Sinova J 2010 *Phys. Rev. Lett.* **105** 227201

[23] Pela R R, Marques M, Ferreira L G, Furthmuller J and Teles L K 2012 *Appl. Phys. Lett.* **100** 202408

[24] Gray A X, Minar J, Ueda S, Stone P R, Yamashita Y, Fujii J, Braun J, Plucinski L, Schneider C M, Panaccione G, Ebert H, Dubon O D, Kobayashi K and Fadley C S 2012 *Nature Mater.* **11** 957

[25] Sadowski J, Domagała J Z, Bąk-Misiuk J, Koleśnik S, Sawicki M, Świątek K, Kanski J, Ilver L and Ström V 2000 *J. Vac. Sci. Technol. B* **18** 1697

[26] Look D C, Walters D C, Manasreh M O, Sizelove J R, Stutz C E and Evans K R 1990 *Phys. Rev. B* **42** 3578

[27] Kuryliszyn-Kudelska I, Domagała J Z, Wojtowicz T, Liu X, Łusakowska E, Dobrowolski W and Furdyna J K 2004 *J. Appl. Phys.* **95** 603

[28] Yastrubchak O, Bak-Misiuk J, Lusakowska E, Kaniewski J, Domagala JZ, Wosinski T, Shalimov A, Reginski K and Kudla A 2003 *Physica B-Condensed Matter*. **340**, 1082

[29] Kochura A V, Aronzon B A, Lisunov K G, Lashkul A V, Sidorenko A A, De Renzi R, Marenkin S F, Alam M, Kuzmenko A P and Lahderanta E 2013 *J. Appl. Phys.* **113** 083905

[30] Potashnik S J, Ku K C, Wang R F, Stone M B, Samarth N, Shiffer P and Chun S H 2003 *J. Appl. Phys.* **93** 6784

[31] Franz W 1958 *Z. Naturforsch.* **13a** 484; Keldysh L V 1958 *Sov. Phys. JETP* **7** 788

[32] Sydor M, Angelo J, Wilson J J, Mitchel W C and Yen M Y 1989 *Phys. Rev. B* **40** 8473

[33] Estrera J P, Duncan W M and Glosser R 1994 *Phys. Rev B* **49** 7281

[34] Aspnes D E and Studna A A 1973 *Phys. Rev. B* **7** 4605

[35] Feenstra R M, Woodall J M and Pettit G D 1993 *Phys. Rev. Lett.* **71** 1176

[36] Szczytko J, Mac W, Twardowski A, Matsukura F and Ohno H 1999 *Phys. Rev. B* **59** 12935

**Table 1.** Layer thicknesses and their lattice parameters, $c$ and $a_{\text{rel}}$, calculated from the results of high-resolution XRD measurements performed at 27°C, Curie temperatures, $T_C$, obtained from SQUID magnetometery measurements, and the values of energies $E_G$ and electro-optic energies $\hbar\theta$, revealed from the analysis of the FKOs periods of the modulation PR spectra, for the investigated epitaxial layers of LT-GaAs, (Ga,Mn)As and (Ga,Mn)As/Sb.

| layer | thickness (nm) | $c$ (Å) (±0.00008) | $a_{\text{rel}}$ (Å) | $T_C$ (K) | $E_G$ (eV) | $\hbar\theta$ (meV) |
|---|---|---|---|---|---|---|
| LT-GaAs, as-grown | 800 | – | – | – | 1.415 | 23.64 |
| LT-GaAs, as-grown | 230 | 5.65789 | 5.65563 | – | 1.413 | 40.61 |
| (Ga,Mn)As, as-grown | 100 | 5.69340 | 5.67294 | 40 | 1.408 | 53.95 |
| (Ga,Mn)As, annealed | 100 | 5.68773 | 5.67018 | 73 | 1.411 | 54.90 |
| (Ga,Mn)As/Sb, as-grown | 100 | 5.69543 | 5.67393 | 50 | 1.408 | 54.60 |
| (Ga,Mn)As/Sb, annealed | 100 | 5.69057 | 5.67156 | 115 | 1.419 | 44.90 |

**Figure captions**

**Figure 1.** High-resolution X-ray diffraction results: 2θ/ω scans for (004) Bragg reflections for LT-GaAs and (Ga,Mn)As (a) and LT-GaAs and (Ga,Mn)As/Sb (b) epitaxial layers grown on (001) semi-insulating GaAs substrate. The narrow line corresponds to reflection from the GaAs substrate and the broader peaks at lower angles are reflections from the layers. The curves have been vertically offset for clarity.

**Figure 2.** SQUID magnetization along the in-plane [$\bar{1}$10] crystallographic direction vs. temperature for the as-grown and annealed (Ga,Mn)As (a) and (Ga,Mn)As/Sb (b) layers after subtraction of diamagnetic contribution from the GaAs substrate. Magnetization hysteresis loops measured at a temperature of 5 K are shown in the insets.

**Figure 3.** Modulation photoreflectance spectra for the two LT-GaAs reference layers of different thicknesses (written in the figure) (a) and for the as-grown and annealed layers of (Ga,Mn)As and (Ga,Mn)As/Sb (b) epitaxially grown on GaAs substrate, where the FKOs extrema are marked with numbers and capital A denotes a contribution related to the layer–substrate interface. The spectra have been normalized to the same intensity and vertically offset for clarity.

**Figure 4.** Analysis of the periods of Franz-Keldysh oscillations revealed in the modulation photoreflectance spectra for the LT-GaAs reference layers and the as-grown and annealed (Ga,Mn)As and (Ga,Mn)As/Sb layers shown in figure 3.

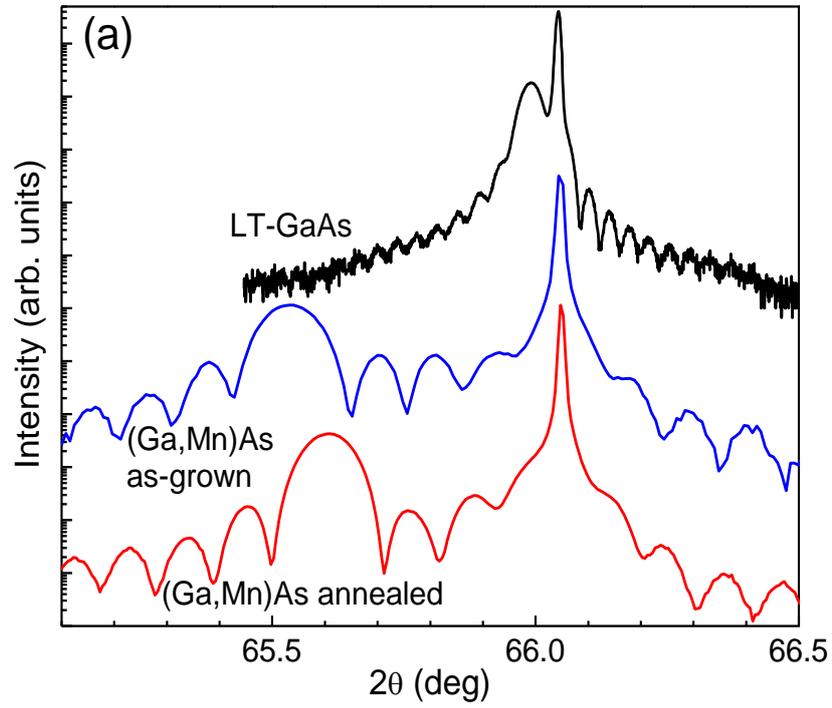

Figure 1a

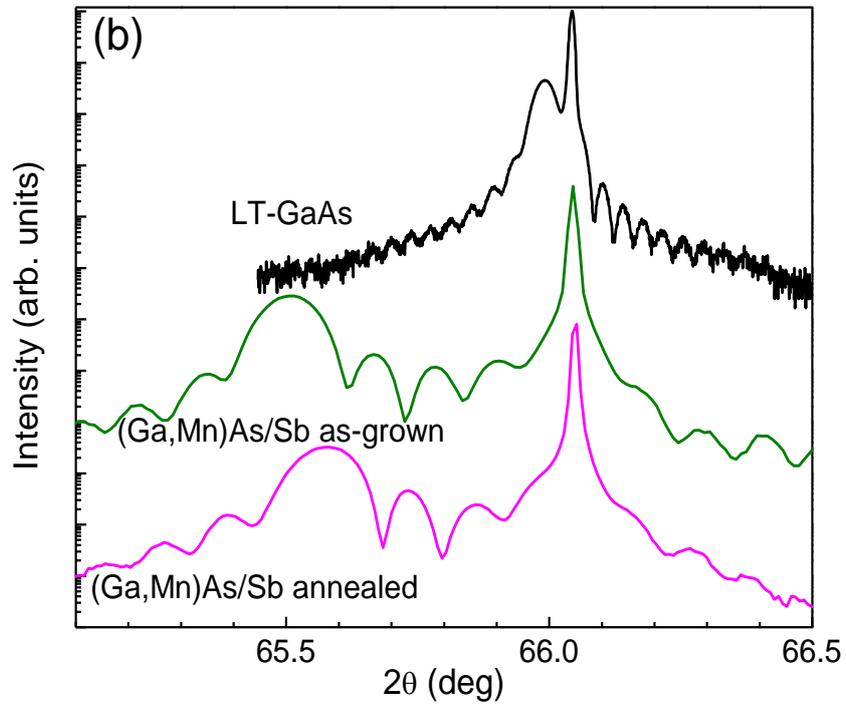

Figure 1b

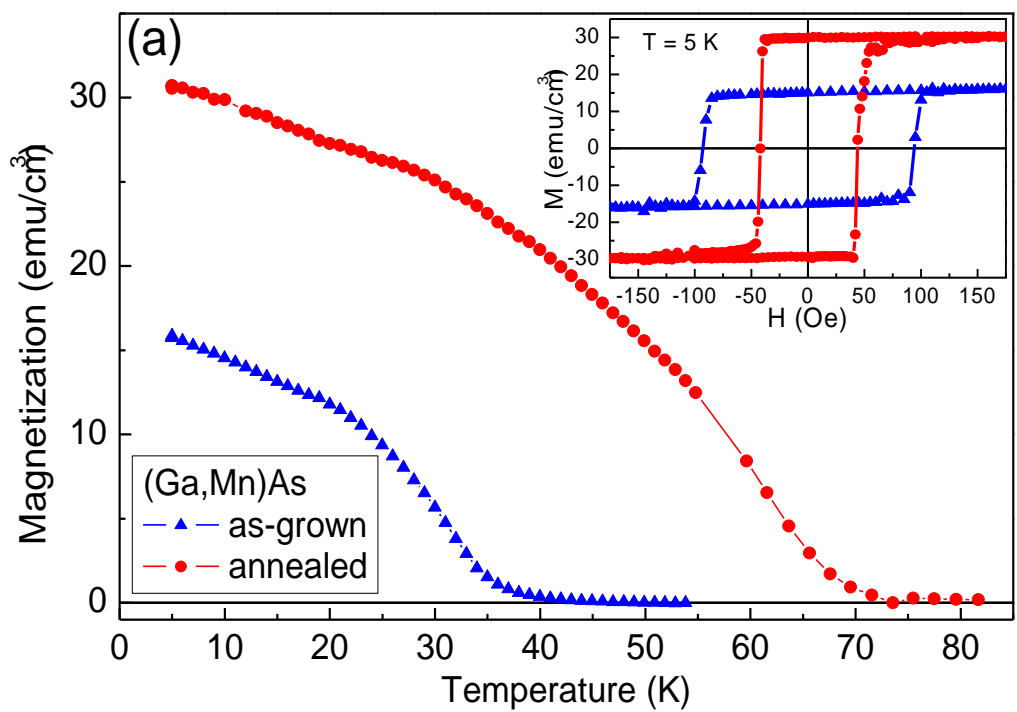

Figure 2a

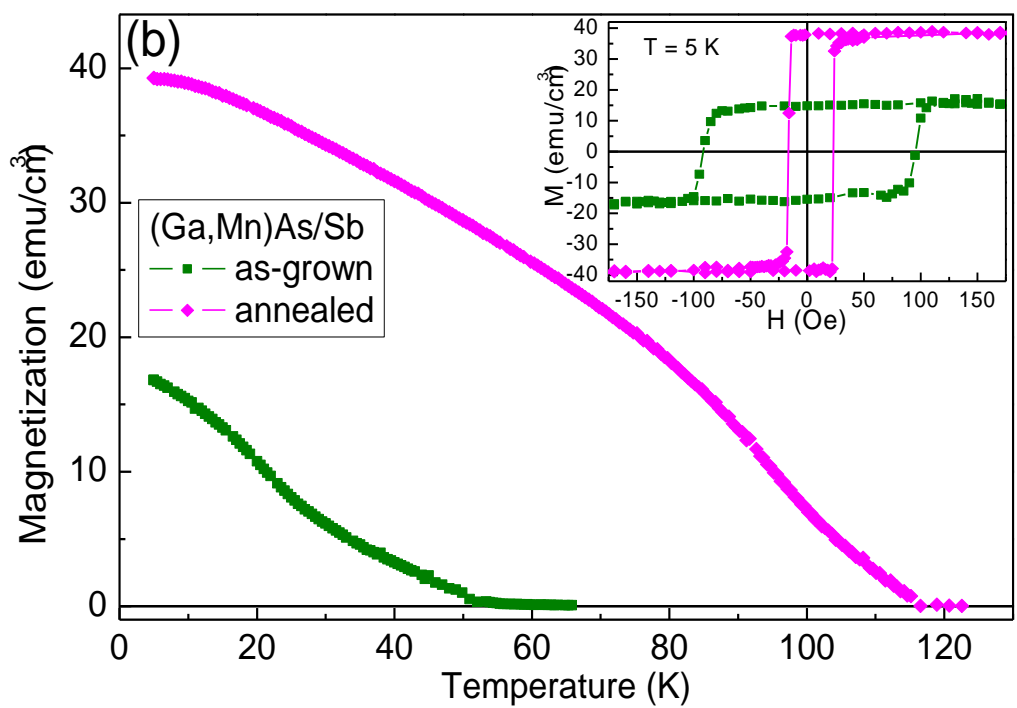

Figure 2b

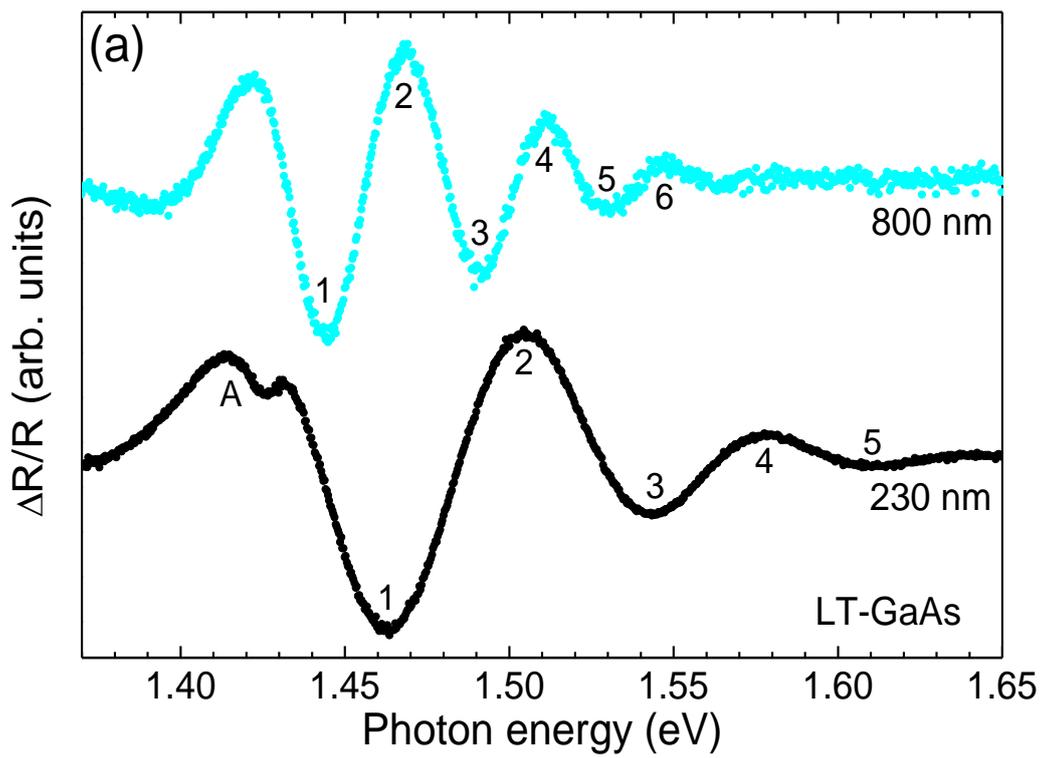

Figure 3a

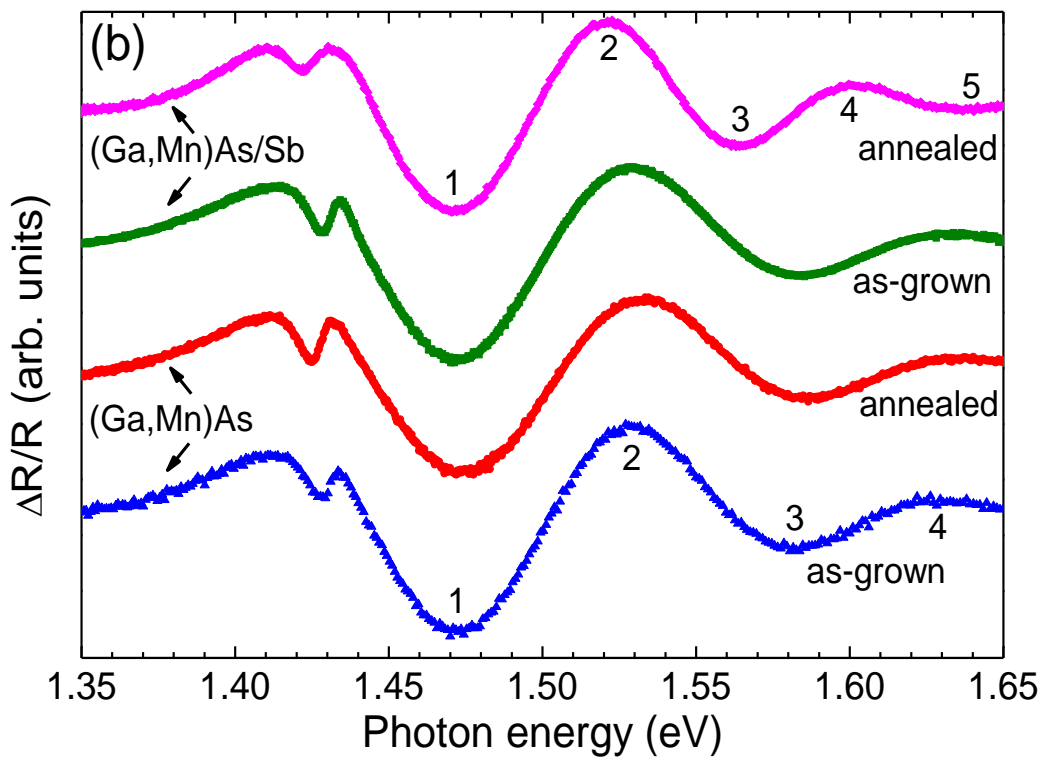

Figure 3b

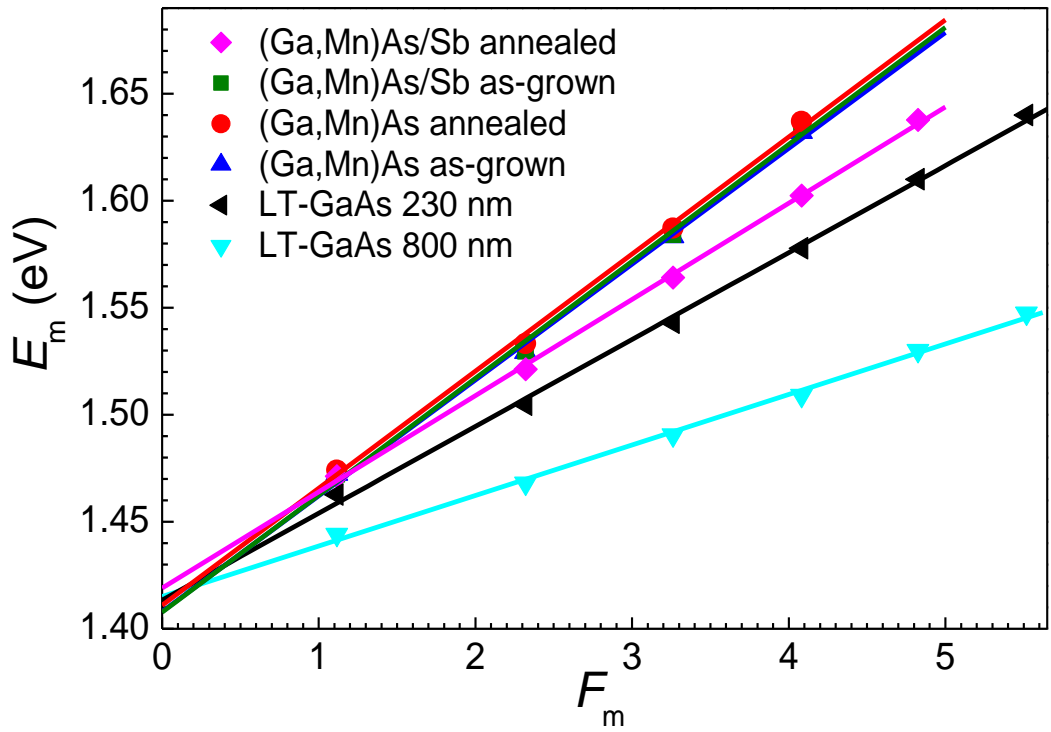

Figure 4